# A room-temperature polymeric spin-valve

Sayani Majumdar[1,2][*], Himadri S. Majumdar[1], Reino Laiho[2] and Ronald Österbacka[1]

[1] Department of Physics, Åbo Akademi University, Åbo 20500, Finland.

[2] Wihuri Physical Laboratory, University of Turku, Turku 20014, Finland.

*Abstract*

We report giant magnetoresistance up to 150% at low bias current and low temperature as well as room temperature magnetoresistance in polymeric spin-valves having the structure $La_{0.67}Sr_{0.33}MnO_3$ (LSMO)/conjugated polymer/Co. The conjugated polymers – regiorandom and regioregular poly(3-hexyl thiophene) were used as the spacer materials. We observed an asymmetric bias voltage dependence of different devices and additional, hitherto unseen, peaks in MR vs. magnetic field plot with low bias currents measurements that we attribute to local magnetic moments due to spin-trapping in the defects in the spacer material. Also, various spacer thicknesses led to variation of magnetoresistance within a certain temperature range.

[*] Corresponding author. E-mail : saymaj@utu.fi    Phone : +358-2-3335942    Fax : +358-2-2154776

Spin electronic ("spintronic") devices, [1] based on the utilization of the spin of electrons, in addition to their charge, opens up an entirely new era of electronics mainly because spins can be manipulated faster and with lower energy than charges [2]. The discoveries of giant magnetoresistance (GMR) [3] and tunneling magnetoresistance [4] in metallic spin valves (SV) have revolutionized applications in magnetic recording and memory. It is envisioned that the merging of electronics, photonics, and magnetics will ultimately lead to new spin-based multifunctional devices and quantum bits for quantum computation and communication. Major challenges for spintronics today are the optimization of electron spin lifetimes, the detection of spin coherence in nanoscale structures, transport of spin-polarized (SP) carriers across relevant length scales and heterostructures, and the manipulation of both electron and nuclear spins at sufficiently fast time scales.

A SV is a layered structure of two ferromagnetic (FM) electrodes with different coercive fields ($H_C$) separated by a nonmagnetic (NM) spacer [5]. During positive to negative sweep of external magnetic field ($B$) the two FM electrodes change their direction of magnetization, changing the relative orientation of the two electrodes between parallel (P) and antiparallel (AP) state. The electrical resistance of the device, therefore, obtains bistable states depending on the scattering of the SP carriers in the electrodes in P/AP alignment. The spacer's role is to decouple the two FM electrodes without hindering the spin transport through it. Several metals and inorganic semiconductors have been investigated to optimize the spin transport. Organic semiconductors, on other hand, composed of much lighter molecules like carbon and hydrogen can give an edge over their inorganic counterparts [6]. They have weak spin-orbit and hyperfine interactions, leading to the possibility of preserving spin-coherence over times and distances much longer than the conventional metals/semiconductors [7]. Recently, Dediu et. al. [8] measured room-temperature SP transport in planar $La_{0.67}Sr_{0.33}MnO_3$(LSMO)/sexithiophene (T6)/LSMO device, and Xiong et al. [9] showed 40% GMR at low temperature (11K) in SVs using an organic semiconductor 8-hydroxyquinoline aluminium ($Alq_3$) as the NM spacer. Petta et al. [10] fabricated magnetic tunnel junctions using a self-assembled monolayer molecular barrier (Ni-octanethiol-Ni) and found 16% MR at 4.2K.

In this article we report results of SVs having LSMO as the bottom and cobalt (Co) as the top electrode and as the NM spacer two commonly used hole-transporting polymers - regio-random and regio-regular poly (3-hexylthiophene) (RRaP3HT and RRP3HT respectively) were used. LSMO is a half-metallic ferromagnet with near 100% spin polarization [11], i.e. it has a fully spin-polarized conduction band. In classical $3d$ magnetic metals (Fe, Ni, Co), the basic layers of GMR systems, the conduction electrons have mainly $4s$ character when the polarized electrons are the $3d$ ones. This electronic structure results in only 40% spin polarization at the Fermi level ($E_F$) for Fe. In LSMO ($T_C$ = 380K) there is no $s$ electron and the band of mainly oxygen $2p$ character lies 2eV below $E_F$ [8] resulting in a fully polarized conduction band of $3d$ character. This motivated the choice of LSMO as the spin-injecting electrode. Furthermore, the possibility of a disordered structure in the polymer-FM transition metal interface [12] is also reduced by the use of LSMO. The choice of the polymer (poly (3-hexylthiophene) (P3HT)) was motivated by several factors. The work function of LSMO and Co is very close to the highest occupied molecular orbital (HOMO) energy of P3HT (5.1eV). So, efficient carrier injection is expected for such half metal-polymer heterostructure. Another advantage of P3HT is the two different polymer morphologies due to different stereo-regular couplings of the side chains. In RRaP3HT films the polymer chains form a one-dimensional structure with entangled chains whereas for RRP3HT films it is a two-dimensional lamellae that lies perpendicular to the substrate [13] leading to much higher conductivity of RRP3HT than that of RRaP3HT [13, 14]. This enables us to study the influence of two completely different electronic structures on spin transport using the same material.

The device structure is shown in the inset of figure 1(a). First, LSMO targets were prepared using standard solid state reaction method [15] and then LSMO films were grown on MgO (100) substrates using pulsed laser deposition at $800^0$C substrate temperature and 0.25 Torr oxygen pressure. All the films were characterized by X-ray diffraction, atomic force microscopy (AFM), zero-field cooled and field-cooled magnetization measurements using SQUID magnetometer down to 5K and resistivity measurements using standard four probe method down to 5K. As found from the X-ray pole figures the LSMO films had c-axis orientation. AFM study shows surface roughness of < 1nm helping to form a good interface in a multilayer structure. After cleaning the LSMO films with acetone P3HT films were spin-coated on top of it from 10 mg/ml chloroform solutions and annealed at $80^0$C for 10 hours in a nitrogen-filled glove-box. Finally Co was vacuum-evaporated on top of the P3HT films and an Al layer was evaporated on top of Co without breaking the vacuum and using the same mask to prevent oxidation of the Co film. The P3HT film thicknesses before the Co deposition varied between 100-200nm. The effective thicknesses of the spacer were smaller due to penetration of Co into the film during evaporation. The MR of the SVs was measured using four probes, one pair of current and voltage leads connected to the LSMO while the other pair to the Al, varying $B$ between +150mT and -150mT and the temperature ($T$) between 5K and 300K.

Current-voltage (*I-V*) curves (Fig. 1(a)) for all the devices with different thicknesses show completely ohmic behavior for $T$ = 5-300K indicating a negligible interface barrier for carrier injection. Due to similar work-functions of LSMO and Co with the HOMO energy of P3HT [16] together with the hole-injection and hole-transport property of LSMO and P3HT, respectively, explains the linearity of the *I-V* curves [8]. Such barrierless heterostructures are very attractive for spin injection in the spintronic applications. The total device resistance ($R_{SV}$) ($\sim 10^3$ - $10^4$ $\Omega$) is several orders of magnitude higher than the in-plane resistivity of the LSMO bottom electrode ($\approx 5.73 \times 10^{-4}$ $\Omega$ cm) eliminating the possibility of any additional geometrical effect. When $T$ is lowered down to 5K $R_{SV}$ shows either a metallic behavior or semiconductor-metal transition (SMT) depending on the NM spacer thickness (Fig. 1(b)), very similar to the LSMO behavior itself except that the SMT temperature of the whole device is much lower compared to the LSMO film. As all the LSMO films are deposited using the same target and same deposition conditions, the SMT temperature and resistance can not vary so widely for different films. We believe that with increasing polymer thickness the device characteristics is shifting from metal to semiconductor like behavior and that explains the lowered SMT temperature and increasing $R_{SV}$. During evaporation, penetration and diffusion of Co in the soft polymer layer also influences $R_{SV}$. However, as $R_{SV}$ is ~ k$\Omega$, we can justifiably predict that inspite of some Co inclusions, there still exists an undoped polymeric layer in which transport is dominated by carrier drift/diffusion. The change in MR can be even higher if the inclusion can be avoided as spin polarization is extremely vulnerable to diffusion of impurities [17]. Investigation on an inclusion-free Co/polymer interface is in progress.

The values of $H_C$ of the LSMO and the Co films were measured at 5K and 300K for the same film thicknesses as used in all the SVs and were found to be 26mT and 10mT for LSMO and Co, respectively [18]. So, when the MR of the devices were measured varying the in-plane $B$ from +150mT to -150mT, the two electrodes become AP between -10mT and -26mT and $R_{SV}$ shows a maximum in this region. For the reverse sweep of $B$ a peak is seen between +10mT and +26mT giving an MR hysteresis loop. MR is defined as $\frac{(R_{SV\max} - R_{SV\min})}{R_{SV\min}}$, i.e. the maximum relative change in $R_{SV}$ within the SV hysteresis. The %MR vs. $B$ in Fig. 2(a) and (b) for RRaP3HT SVs and Fig. 2(e) and (f) for RRP3HT SVs clearly show two peaks while sweeping $B$. We observed 5-50% MR in most RRaP3HT devices at 5K and some devices show even ~160% change in MR depending on the magnitude of external bias current ($I$). The maximum observed MR for RRP3HT devices were about 22% at 5K. For RRaP3HT devices MR increases rapidly with decreasing $I$ while the increase is not that sharp for the RRP3HT devices. The value of minimum $I$, up to which accurate measurements were possible, was restricted by the device resistance. To the best of our knowledge MR>100% has previously been observed only using a semiconducting chalcogenide EuS, which is ferromagnetic below 16.8K, as a spin filter [19].

All our SVs show a normal MR effect ($(R_{SV})_{AP} > (R_{SV})_P$) (see Fig. 2) for all the spacer thicknesses and all the bias currents whereas an inverse MR effect was observed in LSMO/Alq$_3$/Co devices, similar to what was reported by Xiong et al. [9]. LSMO is always positively spin polarized but Co can give positive or negative polarization depending on the modification of the end bonds at the spacer/Co interface [20]. At $E_F$, the Co $d$ band is strongly negative-polarized i.e. the density of states (DOS) of the ↑spin sub-band is less than the ↓spin sub-band DOS. So the negative polarization of Co arises from the selection of $d$-band electrons whereas the positive one shows the selection of the $s$-band electrons [21]. Additionally, Fig. 2(a-f) shows that $R_{SV}$ does not saturate in our devices even after the two electrodes are in P configuration. This is due to the negative high field MR effect of LSMO as shown by Wu et al. [22] for their LSMO/organic diodes.

Three factors – spin injection, spin transport and spin relaxation, determine the efficiency of a SV. Spin injection depends largely on the FM LSMO/polymer interface and the defect sites created there. Smith et al. [23] found that the interfacial atoms deviate from the bulk dimerization, resulting in a small expansion of the end bonds of the LSMO segment and a contraction of the first few polymer bonds. Since LSMO is completely SP at $E_F$, the charge and spin densities coincide in this segment giving us good SP injection.

In Figs. 2(a-f) we have observed an, hitherto unseen, additional $I$ dependent effect in the MR measurements. In the RRaP3HT SVs (Figs. 2(a-d)), as we decrease $I$, the %MR starts to increase and below a certain critical current ($I_C$) two additional peaks appear at $B = 0$ for both increasing and decreasing $B$ and between +10 to +40mT for decreasing $B$ and between -10 to -40mT for increasing $B$. These two additional peaks increase in height with decreasing $I$ and eventually merge so that no hysteresis is visible below $I_C$. We note that this is a consistent device behavior for all the RRaP3HT SVs and can not be negated as noise. Though the physical origin of this behavior is still not very well-understood, it seems to originate from the scattering in the RRaP3HT spacer. Several mechanisms can be responsible for spin relaxation of conduction electrons [24] within the polymer spacer. The *Elliot-Yafet mechanism* suggests that electron spins can relax via momentum scattering if the lattice ions induce spin-orbit coupling in the electron wave function. Momentum scattering is typically caused by impurities (at low $T$) and phonons (at high $T$). The spin relaxation rate being proportional to the momentum scattering rate, electrons suffer more momentum scattering for traveling longer distance and their spins relax faster. We believe that due to small $I$ the SP electrons travel with a much lower drift velocity in RRaP3HT and hence their chance of getting trapped or scattered at interfaces or at the defect sites (Co inclusions also produce additional scattering centers) increases. These trapped polarized spins produce some local magnetic moment in a magnetic field, increasing the inhomogeneity and disorder in the system where more charge carriers can get trapped and give rise to additional higher resistance states. The bias dependence of %MR for RRP3HT SVs (Figs. 2(e-f)) also supports our hypothesis. The %MR increases with decreasing $I$ in this case, but the additional peaks do not appear indicating much less charge/spin scattering in these devices due to higher mobility and better conformational order of RRP3HT [13]. The bias dependence of MR in LSMO/SrTiO$_3$/Co devices [25] showed that DOS of the two FM electrodes at $E_F$ changes with applied bias, thus changing the magnitude and direction of spin polarization changes. The bias dependence of the polymeric SVs are also shown in Fig. 2(g).

However, when the %MR is plotted for different $I$ the RRaP3HT SVs show a maximum close to $I = 0$ while the RRP3HT SVs show a maximum at $I =+80$nA. This implies different modification of the end bonds of the electrode and the spacer at the interface for RRaP3HT and RRP3HT resulting in a shift of DOS at $E_F$ of the FM electrode and change in bias dependence.

There is an asymmetry for forward and reverse bias for all SVs. The %MR decays more slowly with positive bias current than the negative bias current. For RRaP3HT SVs with zero bias, the injecting probability from the LSMO ↑spin $d$ band to the Co ↑spin $d$ band is maximum, thus showing a maximum normal MR. With increasing positive bias, the $E_F$ of LSMO start to go down into energy and injection probability decreases gradually due to smaller DOS at $E_F$ hence showing a slow decrease in SP transport. But with negative current, $E_F$ of LSMO goes up and drastically reduces the ↑spin injection probability and subsequently the %MR. For RRP3HT device at +80nA, the DOS at $E_F$ of LSMO is favorable for maximum ↑spin injection to Co $d$ band showing a maximum %MR at that $I$. For these devices the bias dependence is much less pronounced than the RRaP3HT devices due to its more structured orientation and higher mobility minimizing the spin scattering in the latter devices.

Another striking observation in our experiments is the SV effect in the RRaP3HT devices till quite high temperatures (~ 250K) and in the RRP3HT devices even at room temperature (Fig. 3(a)). Earlier, room temperature SP transport was shown in T6 [8] but our results are the first experimental realization of room temperature MR in polymeric SV devices. The $T$ dependence of %MR for RRaP3HT devices with different thicknesses (Fig. 3(a)) clearly indicates that for a small spacer thickness %MR decreases with *T* showing the effect of decrease of the spin relaxation length in the spacer due to scattering dominated by thermal phonons. But for devices with higher thickness %MR remains constant at low T and then starts decaying. We believe that for higher spacer thickness spin scattering in the device due to interface, impurities and other non-thermal scattering is dominant over phonon scattering and thus the devices have a temperature independent behavior at low T. In the RRaP3HT SVs even the device showing ~160% MR at 5K can not retain the SP transport at 300K but the RRP3HT devices having much lower MR (~22% for *I* =80 nA, ~10% for *I* =5 μA) at 5K can still retain at least some of its spin polarization till 300K and thus shows small but detectable 0.5% MR for *I* =80 nA (Fig. 3(b)).

In conclusion, we have observed GMR (~50%, sometimes 160%) at 5K and room temperature MR in polymeric spin-valves. The asymmetric bias current and temperature dependence of different devices were also discussed. We believe that polymer and organic semiconductors are very good candidates for future spintronic applications.

**Acknowledgement**

Financial support from Academy of Finland Project number 204844 is acknowledged.

**Figure captions:**

**Fig. 1(a).** I-V characteristics at different temperatures for a typical RRaP3HT spin valve showing ohmic behavior. Inset shows the device structure. **(b)** Resistance ($R_{SV}$) vs. temperature ($T$) plot for RRaP3HT (closed symbol) and RRP3HT (open symbol) spin valves. The arrow indicates increasing thickness of RRaP3HT spacer.

**Fig. 2.** %MR vs. magnetic field ($B$) curve for **(a – d)** four different bias currents ($I$) in RRaP3HT SV, showing the appearance of the additional peaks with decreasing current, **(e – f)** for two different bias currents ($I$) in RRP3HT SV, showing absence of any additional peak and **(g)** %MR vs. $I$ plot clearly showing the asymmetry for positive and negative bias currents of a typical RRaP3HT SVs (●) and RRP3HT SVs (○). The inset shows positive bias current dependence of another RRaP3HT SV showing %MR up to 160%. All results from **(a – f)** were for measurements done at 5K.

**Fig. 3(a).** %MR as a function of temperature for the RRaP3HT SVs (●) and RRP3HT SVs (○) measured with 5μA bias current. The arrow indicates increasing thickness of the RRaP3HT spacer. The inset shows one RRaP3HT SV measured with 20 nA bias current. **(b).** Resistance ($R_{SV}$) vs. magnetic field ($B$) plot for the RRP3HT device showing spin-valve effect at room temperature.

**Figures:**

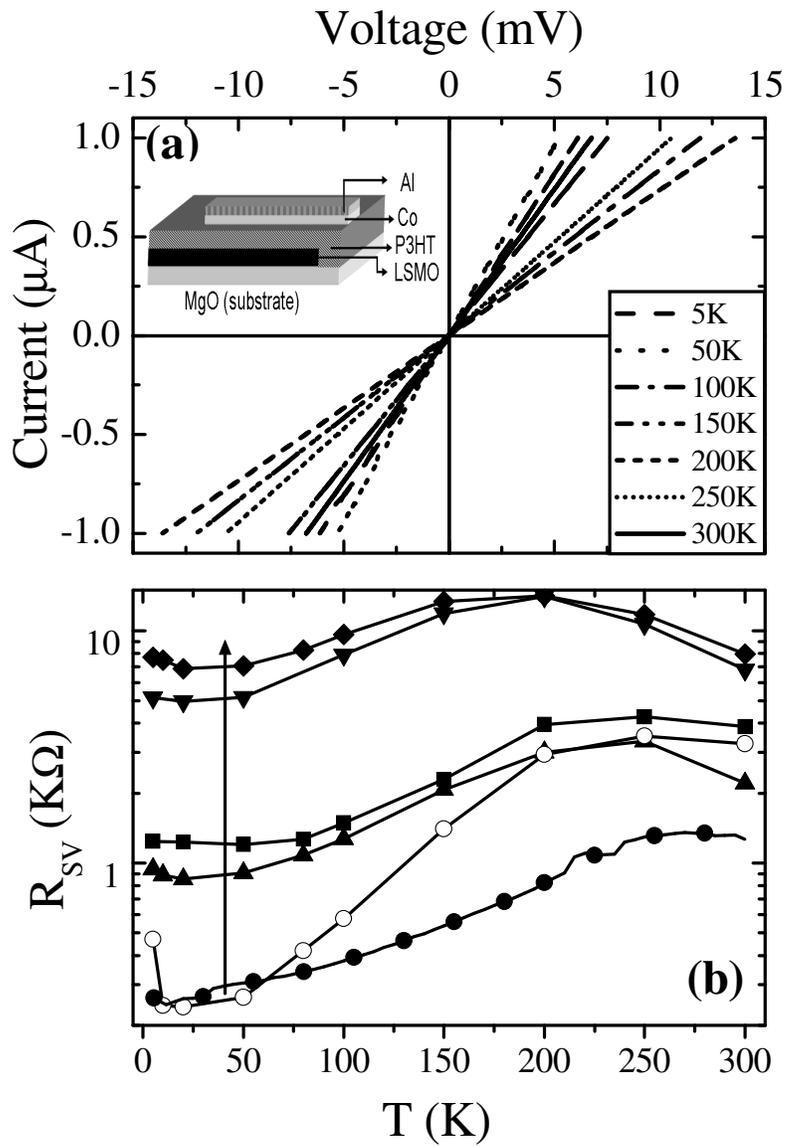

Figure 1. S. Majumdar et. al.

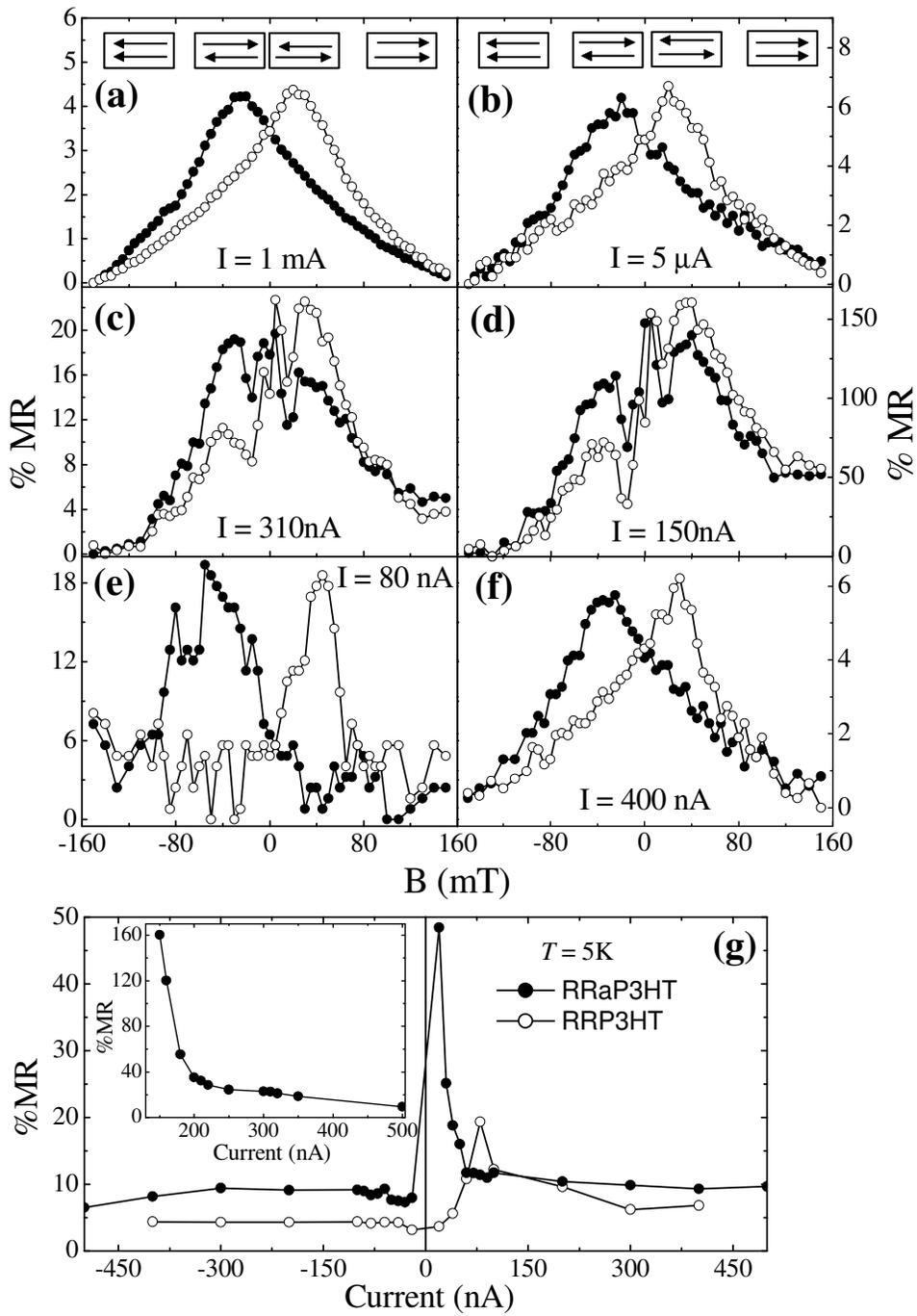

Figure 2. S. Majumdar et. al.

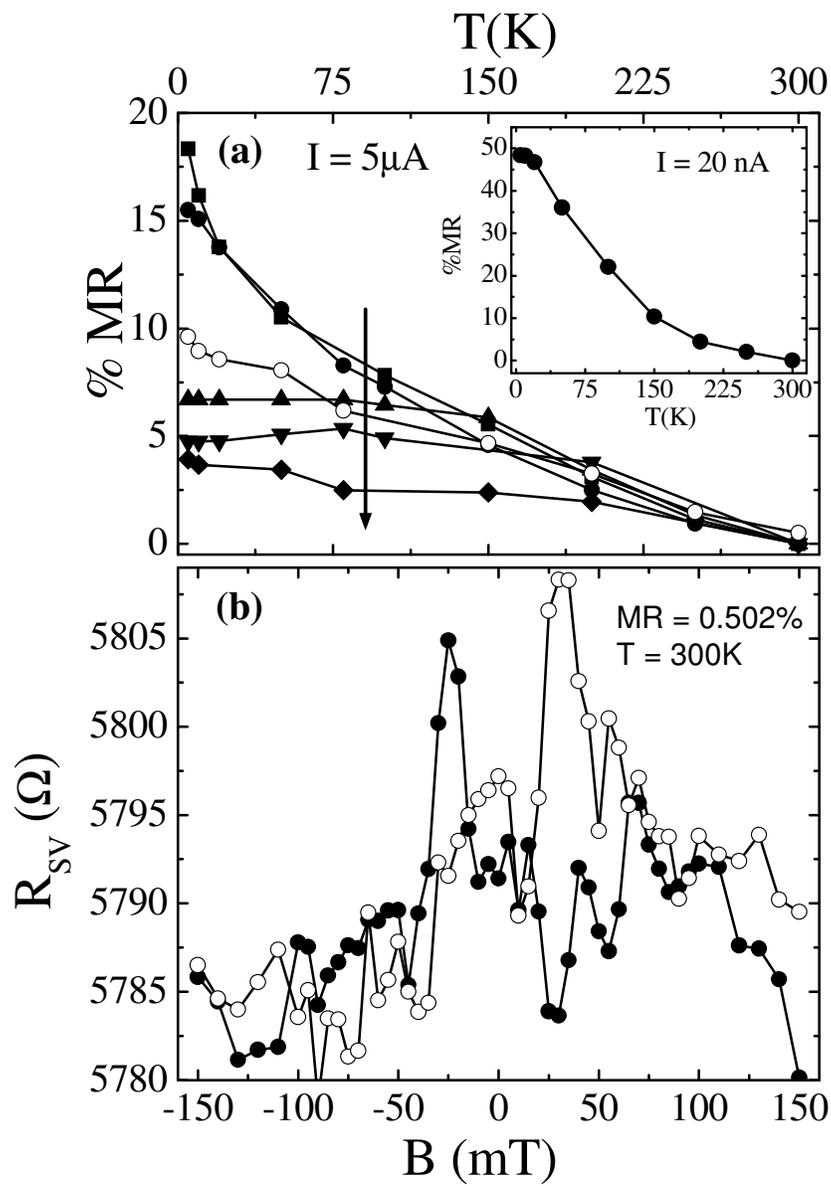

Fig. 3. S. Majumdar et. al.

16